\documentclass[12pt]{elsarticle}

\usepackage{graphics}
\usepackage{amssymb}
\usepackage{mathtext}
\usepackage{nicefrac}

\begin{document}

\begin{frontmatter}

\title{Cosmic ray Nickel nuclei spectrum by the NUCLEON experiment}

\author[dubna,jinr]{V. Grebenyuk}
\author[sinp]{D. Karmanov}
\author[sinp]{I. Kovalev}
\author[sinp]{I. Kudryashov\corref{cor}}
\cortext[cor]{Corresponding author}
\ead{ilya.kudryashov.85@gmail.com}
\author[sinp]{A. Kurganov}
\author[sinp]{A. Panov}
\author[sinp]{D. Podorozhny}
\author[jinr,kiev]{A. Tkachenko}
\author[dubna,jinr]{L. Tkachev}
\author[sinp]{A. Turundaevskiy}
\author[sinp]{O. Vasiliev}
\author[sinp]{A. Voronin}

\address[dubna]{“DUBNA” University, Universitetskaya str., 19, Dubna, Moscow region, 141980, Russia}
\address[jinr]{Joint Institute for Nuclear Research, Dubna, Joliot-Curie, 6, Moscow Region, 141980, Russia}
\address[sinp]{Skobeltsyn Institute of Nuclear Physics, Moscow State University, 1(2), Leninskie Gory, GSP-1, Moscow, 119991, Russia}
\address[kiev]{Bogolyubov Institute for Theoretical Physics, 14-b Metrolohichna Str., Kiev, 03143, Ukraine}

\begin{abstract}

The NUCLEON experiment is designed to measure chemical composition of cosmic rays with charges from $Z=1$ to $30$ in an energy region from $5\cdot10^{11}$ to $10^{15}$~eV. In this article the data analysis algorithm and spectra of Ni and Fe nuclei, measured in the NUCLEON experiment, are presented.

\end{abstract}


\end{frontmatter}

\parindent=0.5 cm

\section{Introduction}

The measurement of the iron peak region of cosmic rays is only a partially solved problem of cosmic ray physics.

This is related to the fact that to measure these nuclei one has to provide a high charge resolution that will allow to separate iron peak elements (from Ti to Ni) from iron itself, as well as a high geometric factor and a wide energy range.

Elements of the iron peak are produced in stellar nucleosynthesis in an equilibrium process (E-process) \citep{intro}, but the lighter than iron elements of the iron peak (Ti, V, Cr) have mainly secondary nature, i.e. they are mostly produced by the fragmentation of iron during propagation of the cosmic ray nuclei in the interstellar medium. Thus elements heavier, than iron, should be measured and compared to iron to investigate the mechanics of the E-processes.

\section{The NUCLEON apparatus}

\begin{figure}[!t]
\begin{center}
\includegraphics*[width=0.9\textwidth]{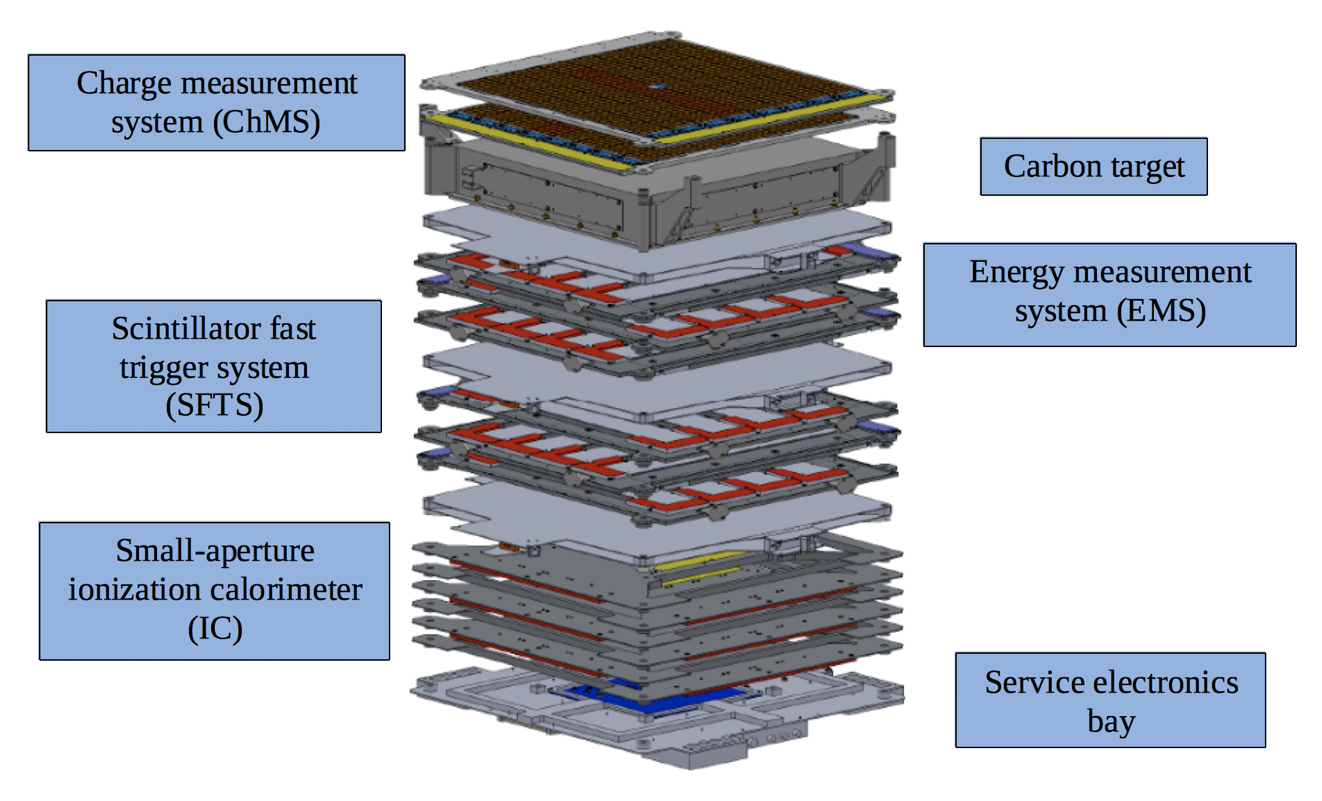}
\end{center}
\caption{Layout of the NUCLEON apparatus.}
\label{nucl}
\end{figure}

The NUCLEON experiment was proposed and developed by Skobeltsyn Institute of Nuclear Physics and Joint Institute for Nuclear Research in Russian Federation.

The NUCLEON apparatus is placed onboard the Russian Resurs-P No. 2 satellite, which was launched to a sun-synchronous circular orbit with average height of 475 km on 26 of December 2014.

The main feature of the apparatus is application of two independent energy measurement techniques – the Kinematic Lightweight Energy Meter (KLEM) method and the traditional calorimetric method.

The KLEM method determines energy of an incident particle by measuring kinematic distribution of fragments of the first inelastic interaction \citep{jcap}.

The traditional calorimetric method determines energy by measuring total energy deposit in the calorimeter, and the implementation will be discussed below.

The NUCLEON apparatus (Fig. \ref{nucl}) consists of:

\begin{itemize}
\item The Charge Measurement System (ChMS) which measures the charge of incident particles. It consists of 4 layers of silicon pad detectors.
\item The Carbon Target (CT) which provides a medium for the first inelastic interaction to occur. It has a thickness of $\nicefrac{1}{4}$ of a nuclear interaction length for protons.
\item The Scintillation Fast Trigger System (SFTS) which provides a trigger for readout electronics. It consists of 3 dual scintillator layers.
\item The Energy Measurement System for the KLEM method (EMS). It consists of 6 layers of silicon strip detectors with a tungsten converter embedded into each plane.
\item A small aperture Ionizing Calorimeter (IC) which is a traditional calorimeter with total thickness of 12 radiation interaction lengths. It consists of 6 planes of silicon strip detectors with a tungsten absorber between them.
\end{itemize}

Total thickness of the NUCLEON apparatus is approx. 14 radiation interaction lengths and 1 nuclear interaction length.

For more details about the apparatus please refer to \citep{nim}.

\section{Data analysis}

To analyze the data the following procedure was used: first the axis of the incident particle was determined, then its charge was determined, and, finally, its energy was determined.

Let’s discuss main features of these steps.

\subsection{Incident particle axis reconstruction.}

\begin{figure}[!t]
\begin{center}
\begin{minipage}{0.55\textwidth}
\begin{center}
\includegraphics*[width=0.9\textwidth]{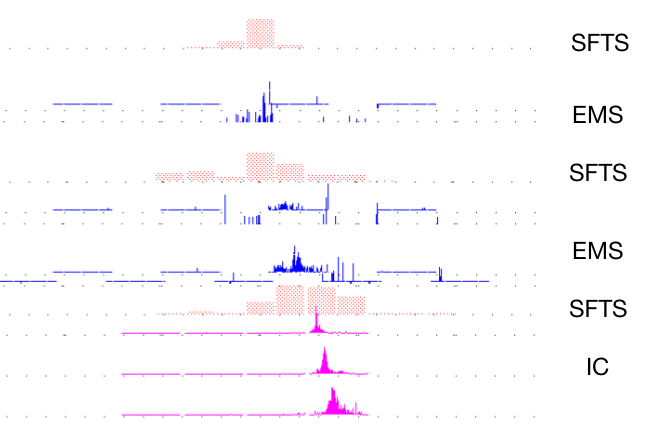}

a.
\end{center}
\end{minipage}
\begin{minipage}{0.35\textwidth}
\begin{center}
\includegraphics*[width=0.9\textwidth]{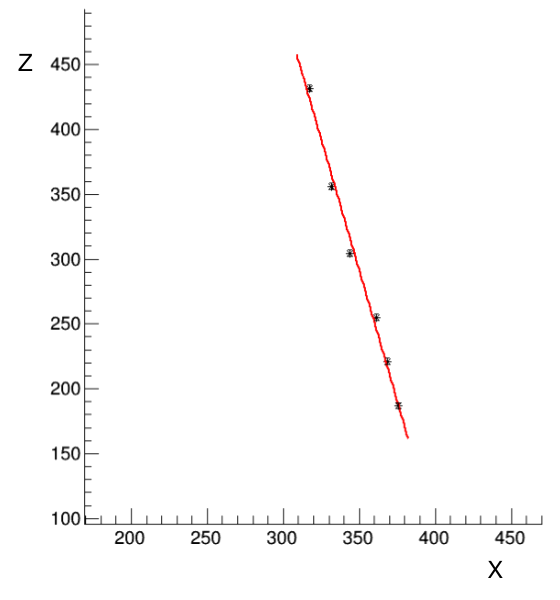}

b.
\end{center}
\end{minipage}
\end{center}
\caption{An example of the axis reconstruction algorithm result – a portrait of one projection (left) and its reconstruction (right)}
\label{axis}
\end{figure}

The reconstruction of the axis was done in four steps:

\begin{enumerate}
\item A maximum energy deposit window with width of 11 mm was found in each layer of EMS and IC.
\item For each window the weighted average of energy distribution, its standard deviation and its error were calculated.
\item Graphs of those weighted averages positions in relation to their coordinates were plotted in two side projections.
\item The graphs were approximated by straight lines which correspond to projections of the primary particle axis (Fig. \ref{axis}.b).
\end{enumerate}

\subsection{Particle charge determination.}

After determining the axis, the next step was charge determination.

Exact identification of the incident particle is required to reconstruct the spectra.

The main difficulty is to separate close parts of CR whose abundance significantly differs, for example, iron peak region nuclei against the background of far more abundant iron.


For each layer of the ChMS a pad with largest signal level was found in a circular region with a radius of 40 mm and its center positioned at the intersection of the plane and the particle’s axis.

A rank statistics, which allows to minimize impact of a non-symmetrical shape of ionization losses in a thin layer, was used on those signals [3].

\begin{figure}[!t]
\begin{center}
\begin{minipage}{0.35\textwidth}
\begin{center}
\includegraphics*[width=0.9\textwidth]{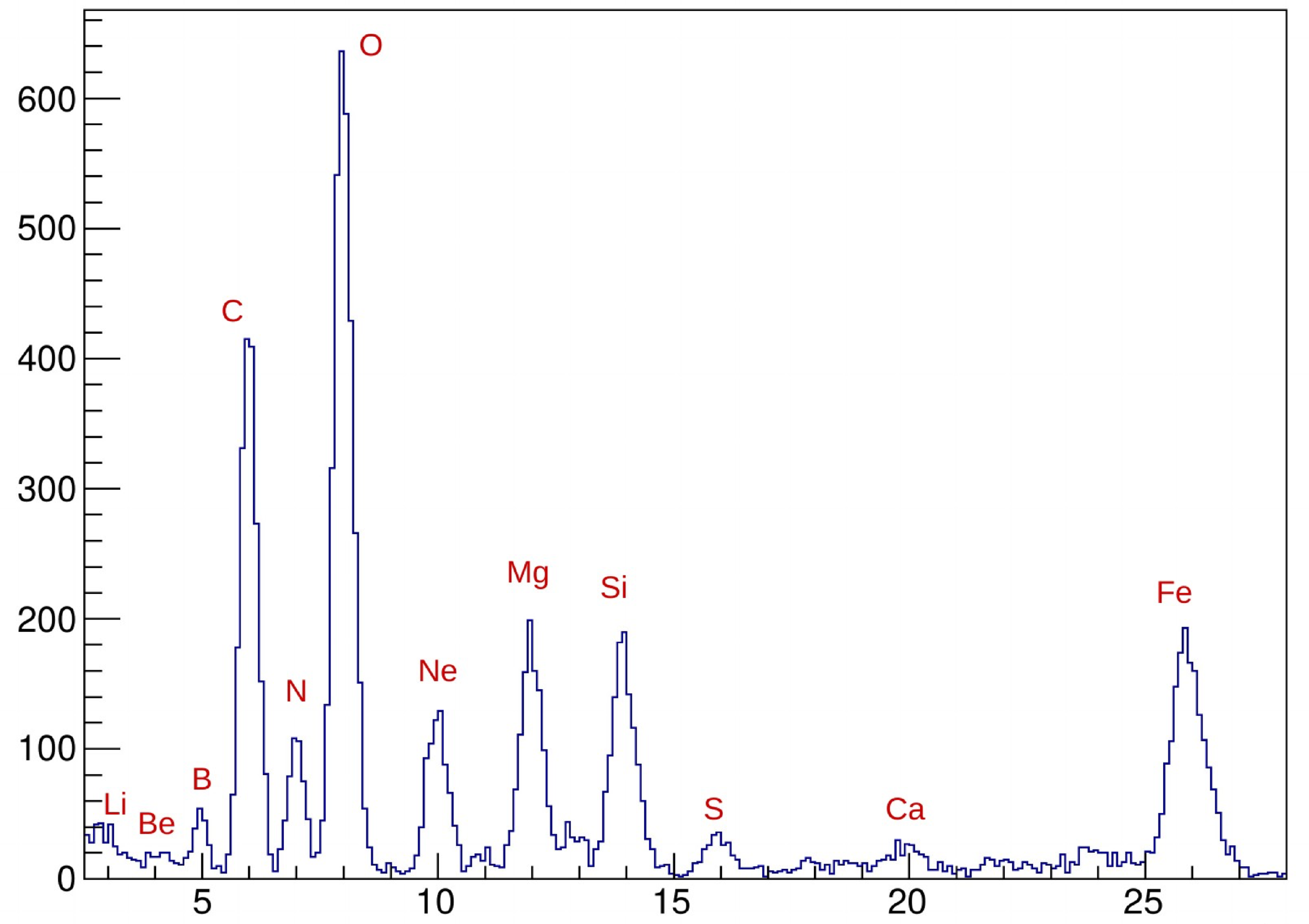}

a.
\end{center}
\end{minipage}
\begin{minipage}{0.55\textwidth}
\begin{center}
\includegraphics*[width=0.9\textwidth]{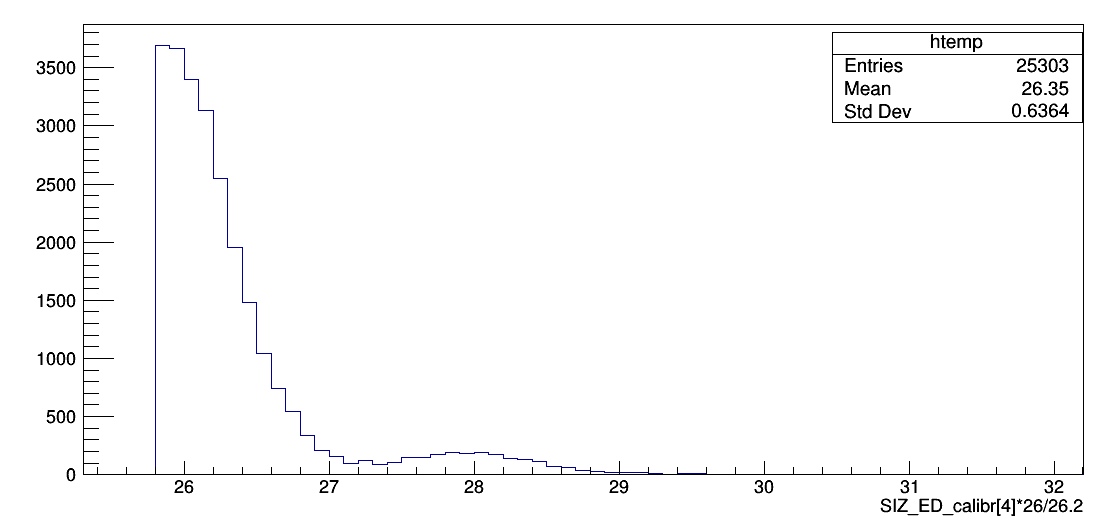}

b.
\end{center}
\end{minipage}
\end{center}
\caption{The charge separation in the NUCLEON experiment.}
\label{chrg2}
\end{figure}

This method allows to confidently separate iron and nickel for events with axis close to vertical ($\sin{\alpha} > 0.85$) (Fig. \ref{chrg2}.b).

\subsection{Energy determination.}

After determining the axis and the charge of a particle its energy was determined.

A new energy measurement method KLEM is described in \citep{jcap}. The technique can be used over a wide range of energies ($10^{11}–10^{16}$ eV) and gives an energy resolution of $70\%$ or better, according to simulation results. We have considerably improved a kinematical method of energy measurements. The simple kinematical method proposed by Castagnoli gives large errors between $100\%$ and $200\%$. A combined method KLEM was proposed to increase resolution. The spatial density of the secondary particles is registered by silicon strip detectors. The estimator calculated as function of the spatial density is applied to reconstruct the primary energy \citep{jcap}.

The total energy deposit of the primary particle in the IC is approximately proportional to the energy of the incident particle. Energy resolution of the IC for hadrons is better than $50\%$ for energies from 100 GeV up to 1 PeV. Low energy resolution is a result of a rather small total length in terms of nuclear interactions, fluctuations of the inelasticity coefficient of the first interaction and large intrinsic fluctuations of nuclear cascades.

The transfer functions of IC are different for each CR component and they are computed separately \citep{jcap}.

\begin{figure}[!t]
\begin{center}
\includegraphics*[width=0.9\textwidth]{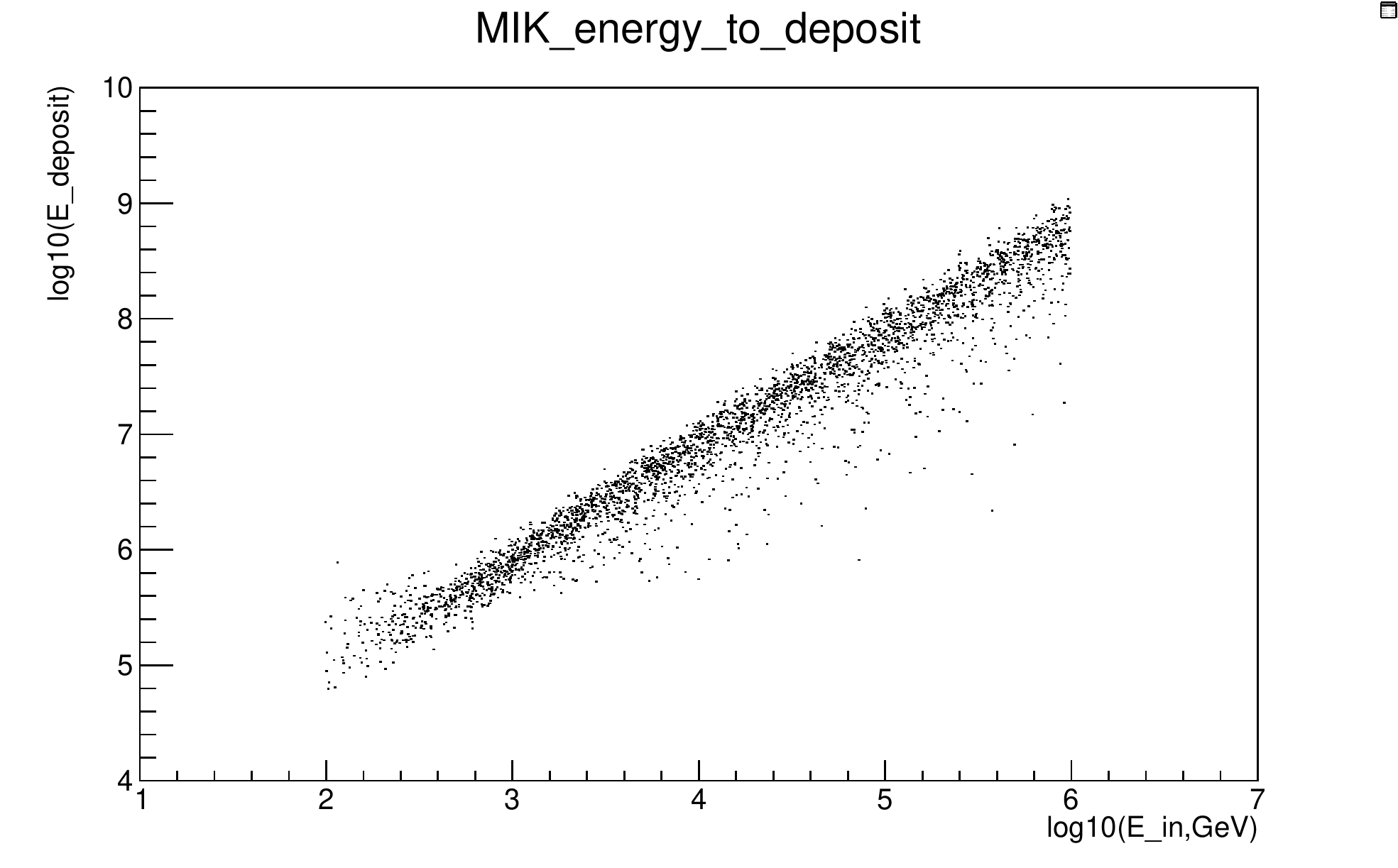}
\end{center}
\caption{The relation between energy deposit and incident energy for iron nuclei.}
\label{rel}
\end{figure}

\begin{figure}[!t]
\begin{center}
\begin{minipage}{0.44\textwidth}
\begin{center}
\includegraphics*[width=0.9\textwidth]{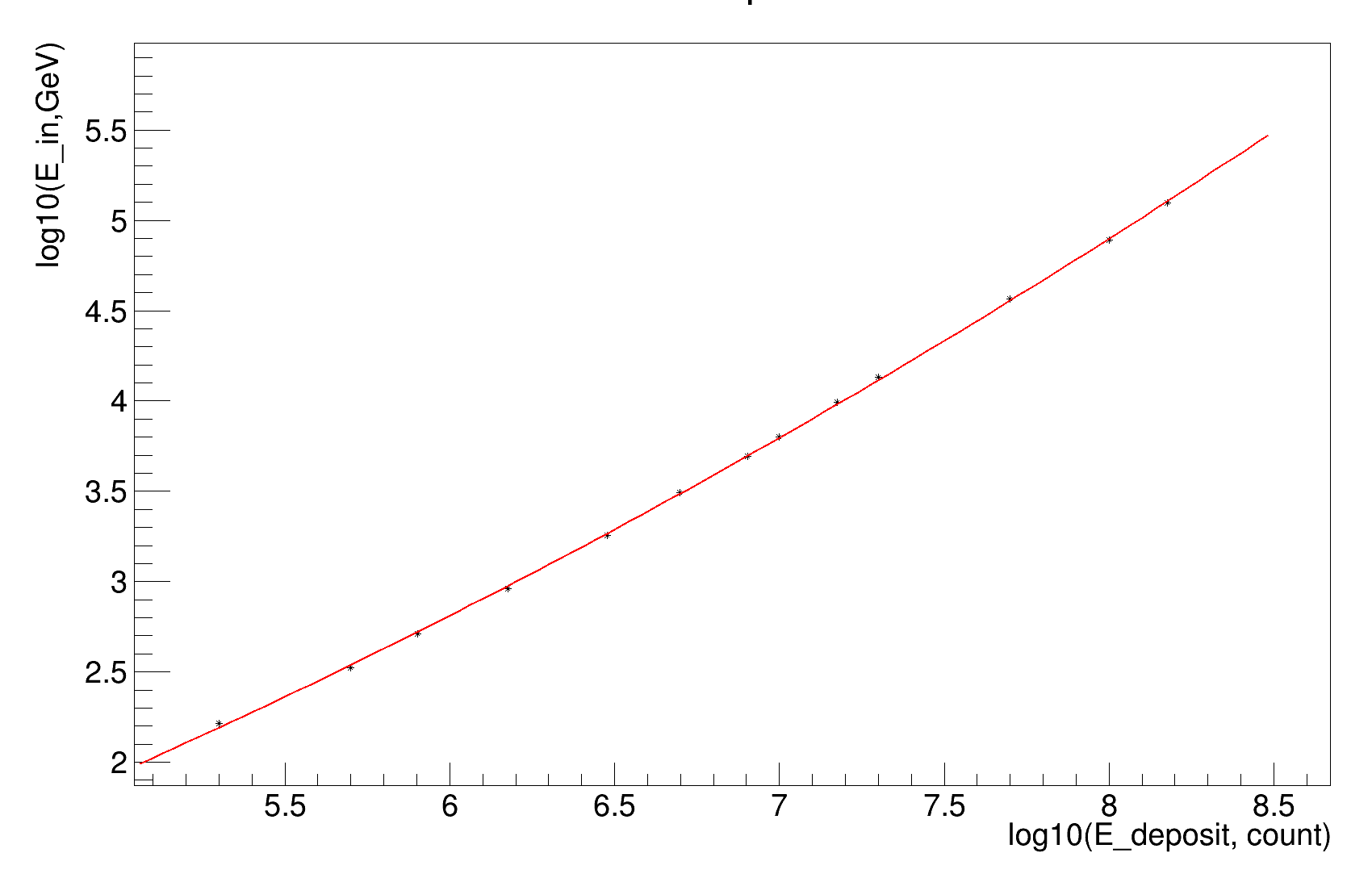}
\end{center}
\end{minipage}
\begin{minipage}{0.46\textwidth}
\begin{center}
\includegraphics*[width=0.9\textwidth]{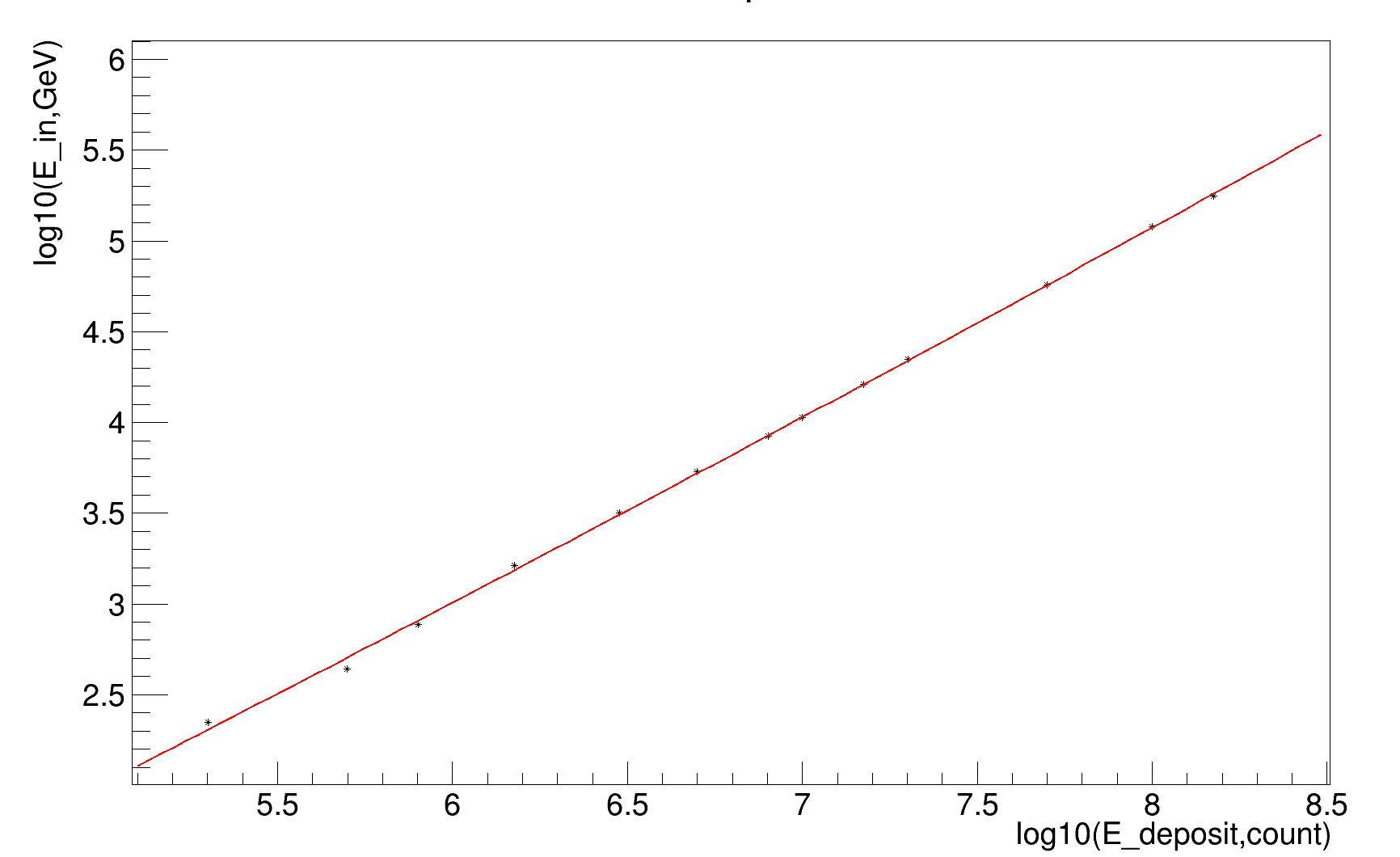}
\end{center}
\end{minipage}
\end{center}
\caption{Relations between incident energy and energy deposit for protons (left) and iron (right), which are approximated by polynomials of degree 2.}
\label{relres}
\end{figure}

Due to high fluctuations of energy deposit and exponential energy dependance of the spectra, a simple approximation of the relation of the energy deposit from the energy of the incident particle would provide a distorted transfer function (Fig. \ref{rel}). Thus, to construct the transfer function for each value of simulated charge a following algorithm was used:

\begin{enumerate}
\item An energy grid was selected.
\item For each value of the grid $\pm10\%$, energy distribution of incident particles was constructed.
\item For each distribution, its mean and standard deviation was determined.
\item A relation between energy deposit in IC and the mean value of the energy of the particle was constructed.
\item The relation was approximated with a polynomial of degree 2 (Fig. \ref{relres}).
\end{enumerate}

As a result, for every simulated nucleus coefficients for a polynomial of degree 2 which binds the energy deposit in the IC to the energy of the incident particle were obtained. For each of these 3 coefficients their dependence of the charge of the particle was constructed. These dependences were approximated by a spline of degree 3.

The transfer functions for every nucleus with charge from 1 to 30 was constructed from these interpolations.

\subsection{Efficiency curve construction.}

To reconstruct the spectra of the cosmic rays, efficiency of the applied algorithms needs to be determined. The efficiency of the trigger system, the axis determination algorithm and the charge determination algorithm needs to be considered.

The Monte-Carlo simulation was used to construct these efficiencies. Because CR spectra are usually presented as binned histograms, efficiency was calculated separately for each such bin.

To achieve this, energy spectra of the simulated particles: as simulated, after the trigger selection, after the axis determination and after the charge determination, were constructed for each component of the CR.

\begin{figure}[!t]
\begin{center}
\begin{minipage}{0.46\textwidth}
\begin{center}
\includegraphics*[width=0.9\textwidth]{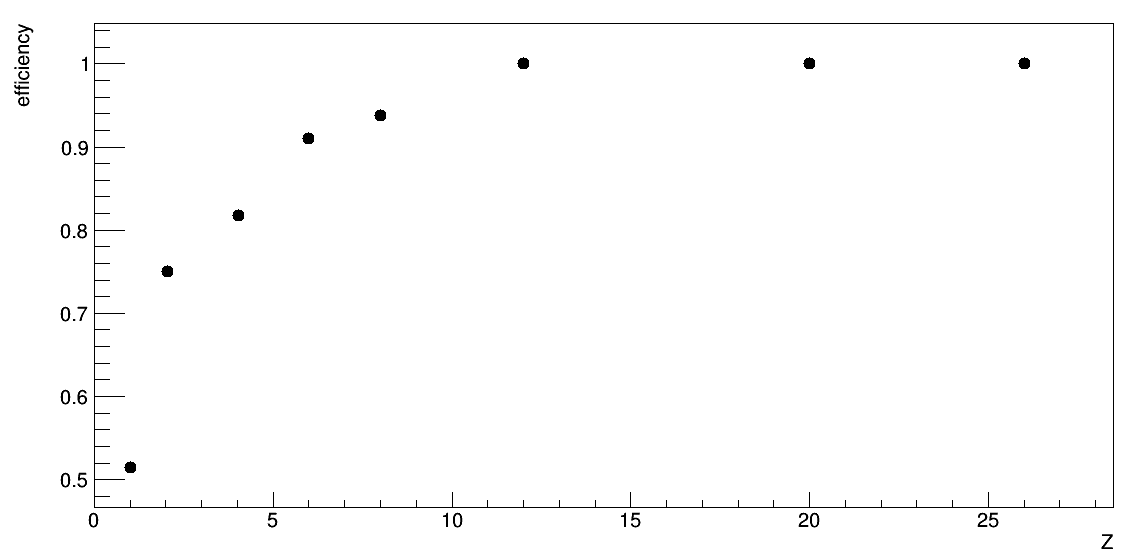}
\caption{The trigger efficiency dependence on the charge Z.}
\label{tref}
\end{center}
\end{minipage}
\begin{minipage}{0.44\textwidth}
\begin{center}
\includegraphics*[width=0.9\textwidth]{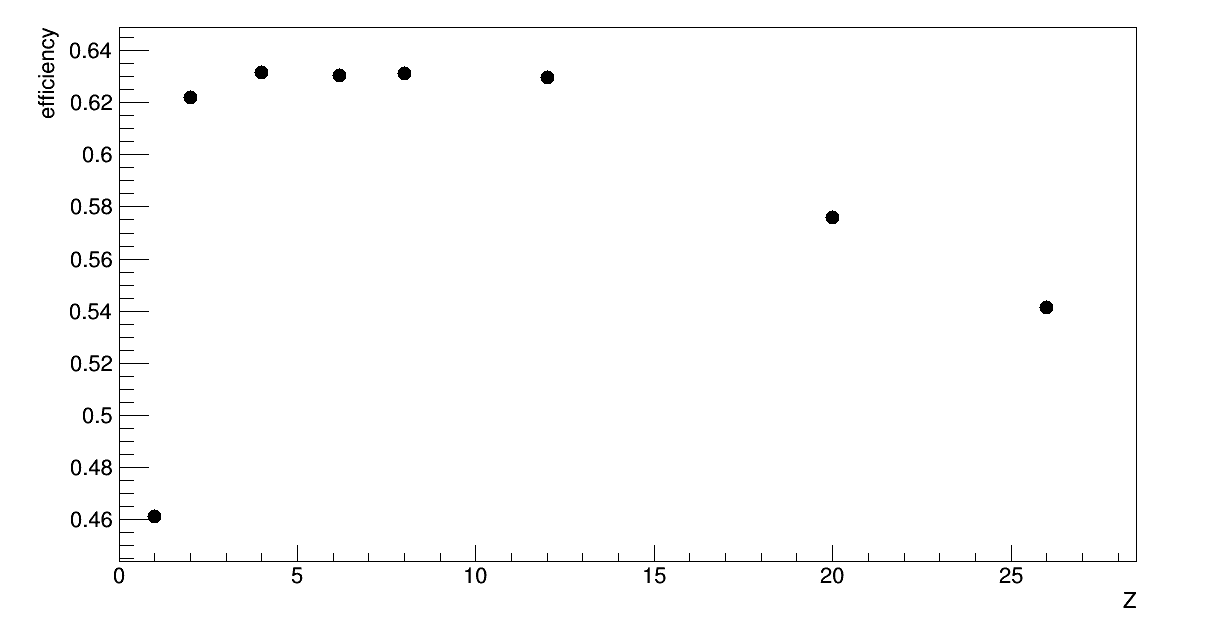}
\caption{The charge reconstruction efficiency dependence on the charge Z.}
\label{chef}
\end{center}
\end{minipage}
\end{center}
\end{figure}

The efficiency of the trigger selection was calculated as a ratio of particle flow after the trigger selection to the flow as simulated for each energy bin of expected spectra (Fig. \ref{tref}). The efficiency of the axis reconstruction algorithm was calculated as a ratio of particle flow with a successfully reconstructed axis to the flow after the trigger selection (Fig. \ref{chef}). The efficiency of charge reconstruction algorithm was calculated as a ratio of particle flow with a successfully reconstructed axis and charge to the flow after the trigger selection. Thus, for each energy bin efficiency of each algorithm was constructed for every simulated charge of the nuclei.

Efficiency values for the rest of nuclei with charges from 1 to 30 were determined by interpolating the relation between the calculated efficiencies and their charge with a polynomial of degree 3.

\section{Spectra}

After processing the flight data through the described algorithms, energy spectra of heavy nuclei, iron and nickel in particular, were obtained.

\begin{figure}[!t]
\begin{center}
\includegraphics*[width=0.8\textwidth]{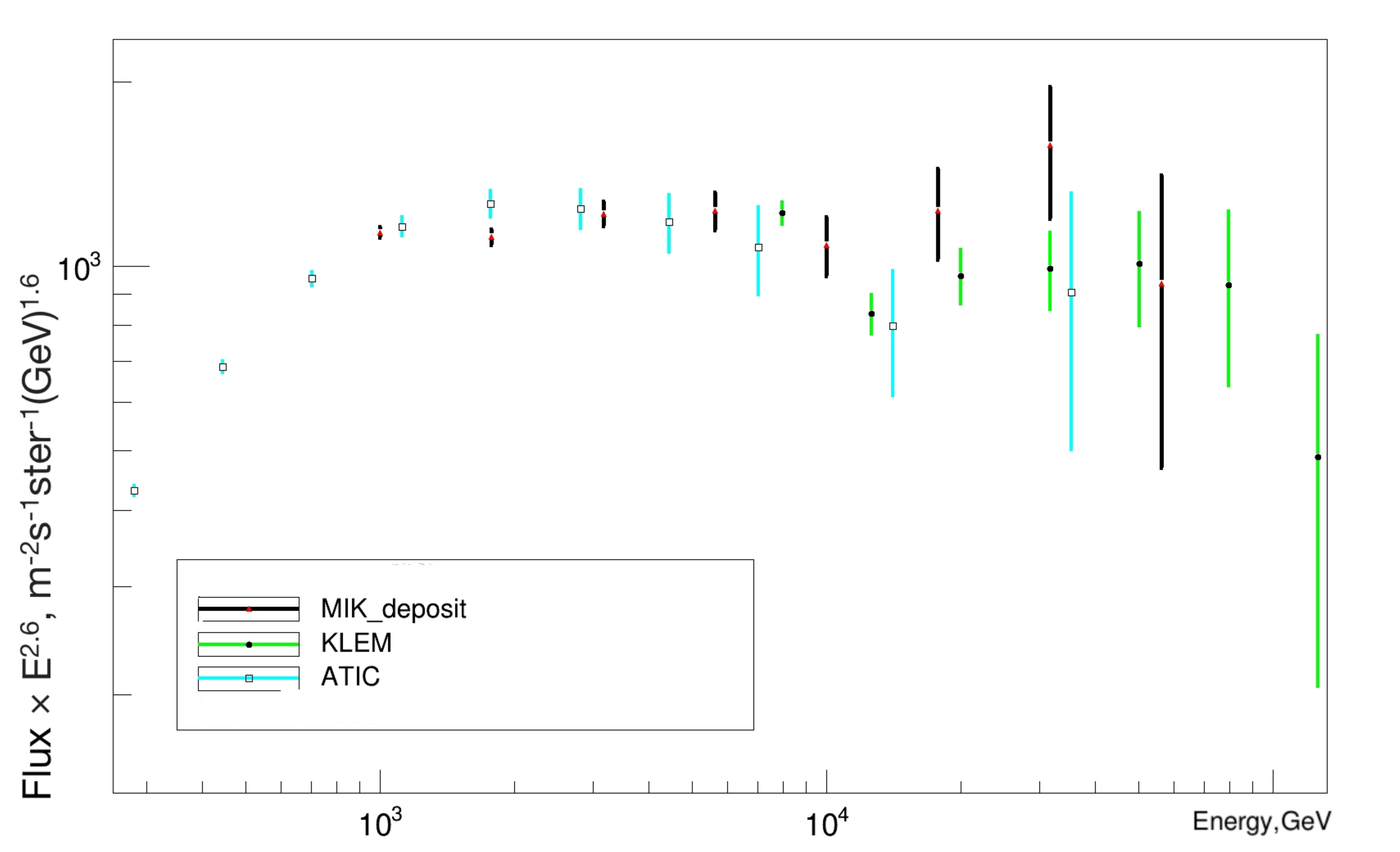}
\end{center}
\caption{The iron spectrum measured by the NUCLEON experiment and the ATIC experiment \citep{atic}.}
\label{fesp}
\end{figure}

\begin{figure}[!t]
\begin{center}
\includegraphics*[width=0.8\textwidth]{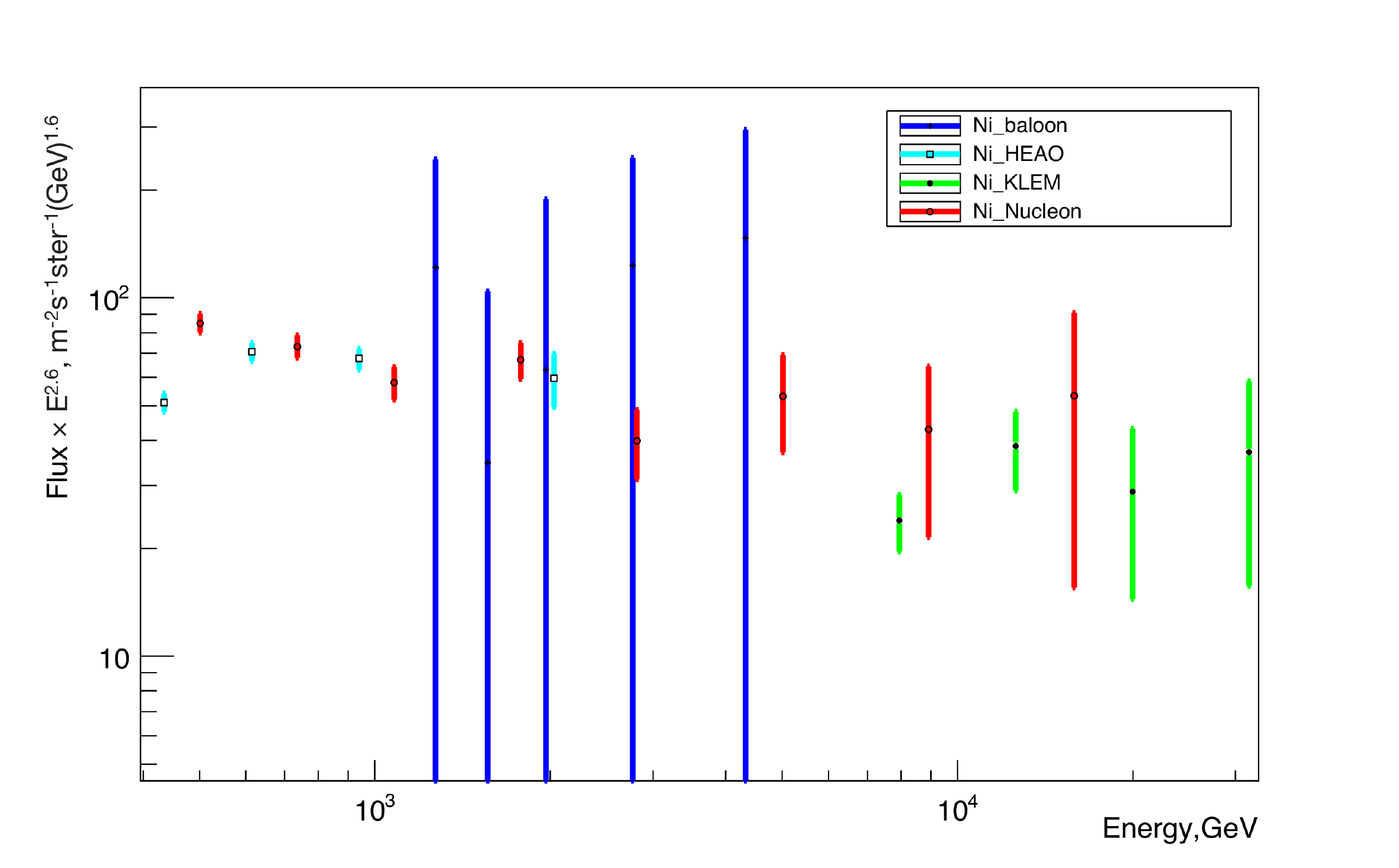}
\end{center}
\caption{The nickel spectrum measured by the NUCLEON experiment, the HEAO-3-C2 experiment \citep{heao} and 1971-1972 balloon experiments \citep{balloon}.}
\label{nisp}
\end{figure}

The iron energy spectrum matches the spectra of other experiments within statistical errors, which confirms that the apparatus was working correctly and the analysis does not contain logical or systematic errors (Fig. \ref{fesp}).

The main result of this article is the nickel spectrum obtained by two independent energy measurement methods (described in this article and \citep{jcap}). The nickel spectrum is obtained in a wide energy region from $10^3$ to $3\cdot10^4$ GeV for the first time (Fig. \ref{nisp}).

Due to nickel abundance being higher than heavier nuclei, the spectrum contains almost no impurities of the fragmentation of heavier nuclei, thus all of the observed nickel was produced by the E-process and accelerated in the CR source.

Iron and nickel nuclei propagation properties are essentially similar due to the small difference in their charges, thus their measured spectra inclination ratio won’t differ much from the source spectra inclination ratio. Therefore, the nickel source spectrum should be substantially softer than the iron spectrum. This observation complements the results obtained by the ATIC collaboration \citep{atic} and can point to fundamental features of CR acceleration in sources.

\section{Conclusions}

The energy spectrum of nickel was obtained for a uniquely wide energy region from $10^3$ to $3\cdot10^4$ GeV from the combined data of the KLEM system and the ionizing calorimeter of the NUCLEON experiment. The inclination of the obtained spectrum is $\gamma_{Ni}=2.83\pm0.09$, which is considerably higher than the inclination of the iron spectrum - $\gamma_{Fe}=2.64\pm0.02$. This result can indicate some peculiarities in the equilibrium process in stellar nucleosynthesis and the Nickel nuclei acceleration.

\section{Acknowledgements}

\end{document}